\documentstyle[12pt, a4]{article}
\begin{document}
\baselineskip = 7 mm
 \begin{center}
\begin{large}
 {\bf {  REEXAMINATION OF STANDARD SOLAR MODEL 
 
 TO THE SOLAR NEUTRINO PROBLEMS }}
\end{large}

\vspace{2cm}

K. FUKASAKU and T. FUJITA 

Department of Physics, Faculty of Science and Technology  
  
Nihon University, Tokyo, Japan 

\vspace{3cm}

{\large ABSTRACT} 

\end{center}

We examine the calculation of the solar neutrino flux based on the 
standard solar model (SSM). It is found that the solar neutrino data 
( KAMIOKANDE experiment ) 
can be well described by the SSM with careful employment of 
nuclear data of $^7Be (p,\gamma ) ^{8} B$.  The  main point is 
that the simple-minded product ansatz of Coulomb plus nuclear parts 
should have a few percent uncertainties which induce the large 
reduction of the neutrino flux from $^8 B$. Also, if the electron 
capture of $^7 Be$ inside the sun is suppressed, then the GALLEX 
experiment can be understood by the SSM calculation.

\newpage
 
\begin{enumerate}

\item{\large Introduction}

The solar neutrino problem is a long standing puzzle. The discrepancy 
between theoretical predictions of the neutrino flux 
by the standard solar model (SSM) and the observed data is still 
believed to be a factor of 2 or more [1].  This problem, however, has produced 
many different kinds of refinements of solar internal structure model 
as well as new ideas in neutrino physics such as neutrino oscillations [2]. 

In this paper, we reexamine the calculation of the standard solar model 
by carefully considering the nuclear reaction data. To this claim, we may 
face criticisms that the nuclear reaction parts must have already 
been examined very carefully by all of the previous calculations.  
This is certainly right. 
The nuclear reaction data have been improved a lot and only those refined 
data have been employed. 

However, there is one important point which is required to reconsider
 in the previous calculations. 
That is, the Coulomb part calculated by the WKB method. The Coulomb 
coefficients can be calculated quite reliably 
if it is only one body problem.  However, if it involves many body 
nature in the nuclear reaction, it is not very clear to what accuracy 
one can believe the WKB results even though we know that they cannot be 
very bad. 

Also, one knows in nuclear physics that the Coulomb problem is 
not as simple as one at first thinks. The Nolen-Schiffer anomaly 
is a good example [3]. The Coulomb displacement energy is not well described 
if one wants to discuss it to a very high accuracy [4]. 

Here, the problem is that the high energy neutrino flux is 
very sensitive to the Coulomb coefficients. In fact, the few percent 
change of the Coulomb coefficients may sometimes induce a large effect on the 
neutrino flux, leaving most of the solar structure quantities unchanged. 
In particular, the nuclear reaction data of $^7Be (p,\gamma ) ^{8} B$ is 
most sensitive to the high energy part of the solar neutrino flux. 
As we will see below, a few percent increase of the Coulomb coefficient 
is enough to reduce the neutrino flux by a factor of 5. Furthermore, 
the choice of the new Coulomb coefficient is perfectly consistent with 
the existing reaction data of $^7Be (p,\gamma ) ^{8} B$ [5]. 
At the same time, we 
reproduce all of the physical quantities of the solar internal 
structure at the same level of accuracy as the previous calculations of SSM. 

To summarize our results of the neutrino flux, we obtain the following 
neutrino capture rates for GALLEX [6], KAMIOKANDE [7], SAGE [8] and 
Homestake (Davis et al. [9] ). Here, BP95 and DS96 denote the recent 
calculations by Bahcall and Pinsonneaul [10], 
and Dar and Shaviv [11], respectively. 
We present the two different calculations ( Case I and Case II ) which 
will be explained below in detail.

\vspace{1cm}
\newcounter{NO}\newcommand{\NO}[1]{\setcounter{NO}{#1}\Roman{NO}}

\begin{center}

\underline{Neutrino Flux}\\
\vspace*{0.5cm}
 
\begin{tabular}{c|c|c|c|c|c}
\hline
 & \multicolumn{2}{c|}{\raisebox{-0.2ex}[0pt]{Present cal.}}  &&&  \\
\cline{2-3}
 & \raisebox{-0.2ex}[0pt]{\makebox[10mm]{I}}
 & \raisebox{-0.2ex}[0pt]{\makebox[10mm]{II}} 
 & \raisebox{1ex}[0pt]{BP95} & \raisebox{1ex}[0pt]{DS96}
 & \raisebox{1ex}[0pt]{Experiment}  \\
\hline
\hline
&&&&& \\
   \raisebox{1.3ex}[0pt]{Homestake(SNU)} 
 & \raisebox{1.3ex}[0pt]{4.5} 
 & \raisebox{1.3ex}[0pt]{3.4} 
 & \raisebox{1.3ex}[0pt]{$9.3\pm1.4$} 
 & \raisebox{1.3ex}[0pt]{$4.1\pm1.2$} 
 & \raisebox{1.3ex}[0pt]{$2.55\pm0.17\pm0.18$} \\
\hline
   \raisebox{-0.3ex}[0pt]{KAMIOKANDE} &&&&& \\

   \raisebox{0.3ex}[0pt]{(10$^{6}$cm$^{-2}$sec$^{-1}$)}
 & \raisebox{1.3ex}[0pt]{2.9} 
 & \raisebox{1.3ex}[0pt]{1.9} 
 & \raisebox{1.3ex}[0pt]{6.62} 
 & \raisebox{1.3ex}[0pt]{2.49}
 & \raisebox{1.3ex}[0pt]{2.73$\pm$0.17$\pm$0.34} \\
\hline
&&&&& \\
   \raisebox{1.3ex}[0pt]{GALLEX(SNU)} 
 & \raisebox{1.3ex}[0pt]{116} 
 & \raisebox{1.3ex}[0pt]{114} 
 & \raisebox{1.3ex}[0pt]{137$\pm$8} 
 & \raisebox{1.3ex}[0pt]{115$\pm$6} 
 & \raisebox{1.3ex}[0pt]{77.1$\pm$8.5$^{+4.4}_{-5.4}$} \\
\hline
&&&&& \\
   \raisebox{1.3ex}[0pt]{SAGE(SNU)} 
 & \raisebox{1.3ex}[0pt]{116}
 & \raisebox{1.3ex}[0pt]{114}
 & \raisebox{1.3ex}[0pt]{137$\pm$8}
 & \raisebox{1.3ex}[0pt]{115$\pm$6}
 & \raisebox{1.3ex}[0pt]{69$\pm$10$^{+5}_{-7}$} \\
\hline 
\end{tabular}
\end{center}

\vspace{3cm}

\item{\large The Standard Solar Model}

The internal structure of the sun is by now described reasonably well  
by the standard solar model. The chain of nuclear reactions is well 
understood. The description of the sun reduces to several couples of 
differential equations which should be solved mostly by numerical 
calculations.  
Among the parameters that enter in the equations, the opacity 
coefficient must be most ambiguous. However, recent studies to refine  
the SSM enable us to remove the ambiguity of the opacity fairly well. 
In connection with the solar neutrino problems, the ambiguity of the opacity 
may lead to a correction of a few tens of percents to the neutrino flux. 
In this respect, we have only a very little freedom left for neutrino 
flux [1].  

The energy of the sun is governed by the nuclear reaction cross sections. 
The energy production rate $\epsilon_{12}$ for $1+2 \rightarrow 3+4+Q$ 
reaction is described by 

$$ \epsilon_{12} = {QN_1 N_2 < \sigma v >  
\over{ \left( 1+\delta_{12} \right) \rho}} 
\quad {\rm erg/g} \cdot {\rm s} $$

where $N_1$ and $N_2$ are the number of particles in the reactions. 

Further, the $ < \sigma v > $ can be parametrized for nuclear 
reactions in the following way
 except $^7 Be ( e,\nu_e ) ^7Li $ reaction, 

$$ N_A < \sigma v > = C_1 T_9^{-{2\over 3}} \exp \left( -C_2T_9^{-{1\over 3}} 
- \left( {T_9\over{T_0}} \right)^2 \right)   $$ 
$$ \times 
\left( 1+C_3T_9^{{1\over 3}}+C_4T_9^{{2\over 3}} +C_5T_9+C_6T_9^{{4\over 3}} +
C_7T_9^{{5\over 3}} \right) +C_8T_9^{-{2\over 3}} 
\exp \left( -C_9T_9^{-1} \right)  \eqno{(2.1)}  $$

For the $^7 Be ( e,\nu_e ) ^7Li $ reaction, we employ the following 
form,
$$ N_A < \sigma v > = 1.34\times 10^{10} T_9^{-{1\over 2}} $$
$$ \qquad \times  
 \left(1 -0.537T_9^{{1\over 3}} 
+3.86 {T_9^{2\over3}}+1.2T_9+0.0027T_9 
\exp \left( {0.002515\over{T_9}} \right) \right)  \eqno{(2.2)} $$

Here, $N_A$ denotes Avogadro number. $T_9$ is measured by $10^9$ $K$. 

The values of the parameters $C_1,..., C_9, T_0 $ are determined from the 
nuclear reaction data and are listed in ref.[1,12]. 

The temperature $T$ and the 
density $\rho$ of the sun are determined by solving the following 
coupled equations, 

$$ {dP\over{dr}}= - {GM\rho \over{r^2}}   \eqno{(2.3a)}  $$

$$ {dM\over{dr}}=  4\pi r^2 \rho    \eqno{(2.3b)}  $$

$$ {dL\over{dr}}=  4\pi r^2 \rho \epsilon   \eqno{(2.3c)}  $$

$$ {dT\over{dr}}= - {3\kappa\rho L \over{16\pi acr^2 T^3}} \qquad   
{\rm for \ \  radiative} \eqno{(2.3d)}  $$

$$ {dT\over{dr}}= {1\over{(n+1)_{ad}}}{T\over{P}}{dP \over{dr}} \qquad 
 {\rm for \ \  convective}  \eqno{(2.3e)}  $$

where $P$, $M$ and $L$ denote the pressure, the interior mass and 
the luminosity, respectively. Also, $a$ and $c$ are radiation density 
constant and the velocity of light. Further, $\kappa$ and $(n+1)_{ad}$ 
denote the opacity and adiabatic coefficient, respectively. 

\vspace{3cm}

\item{\large $^7Be (p,\gamma ) ^{8} B$ reaction}

Now, we want to discuss the nuclear reaction $^7Be (p,\gamma ) ^{8} B$ 
since this is obviously the most important reaction to produce high 
energy neutrinos. In particular, we want to focus on the penetration 
factor $P_{Coul}$. This is expressed in terms of the WKB calculation as 

$$ P_{Coul} ={C_0\over{\sqrt{E}}} \exp \left( -2\pi \eta \right)
  \eqno{(3.1)} $$

where $C_0$ is a constant and $\eta$ can be described as 

$$ \eta = {Z_1 Z_2 e^{2}\over{\hbar v}} = 
{Z_1 Z_2 e^2 \sqrt{\mu}
\over{\hbar \sqrt{2} }}{1\over{\sqrt{E}}}  \eqno{(3.2)} $$
where $\mu$ denotes the reduced mass of the interacting particles. 
To the first order approximation, one may assume that 
the cross section can be described as a product of 
$P_{Coul}$ and the nuclear part $P_{Nucl}$ 
that is connected to the probability 
to make  nuclear reactions. 

$$ \sigma (E) = {1\over{v}} P_{Coul} P_{Nucl} = {S(E)\over E} 
  \exp (-2\pi \eta)  \eqno{(3.3)} $$

where $S(E)$ is a nuclear spectroscopic factor. 

Now, the question is to what accuracy we can believe 
the product ansatz of eq.(3.3) even though the WKB estimation is taken to 
be reliable.  This is connected to the fact that 
the nuclear reaction of  $^7Be (p,\gamma ) ^{8} B$ should be treated 
as a many body problem. Recent calculations by Brown et al.[13] show rather 
a large value of $S(0)$. 

On the other hand, Xu et al. [14] claim that the $S(0)$ value extracted from 
$^8 B \rightarrow p+Be$ decay vertex constant is consistent with 
the observed value of Filippone et al. ($S(0) \sim 17.5 \   eV$ ) [5]. 
Thus, it is still far beyond determining the 
Coulomb coefficient to the accuracy of a few percent in the realistic 
nuclear many body calculations. 

Here, we do not want to rely on the simple-minded product ansatz of eq.(3.3). 
Instead, we assume the following 
form for $\sigma (E)$, 

$$ \sigma (E) = {B\over{E}} \exp \left[ 
 {-{A\over{\sqrt{E}}}} \right]  \eqno{(3.4)} $$ 

where $A$ and $B$ are free parameters which should be determined by 
reproducing the nuclear reaction data.  We stress that our aim is not to 
reproduce theoretically the cross section data, but to find out some 
parameter sets that reproduce the observed cross section [5].   

In fig.1, we show the comparison of the observed cross section of  
 $^7Be (p,\gamma ) ^{8} B$ with that reproduced  
by eq.(3.4) with two choices of the parameter set $A$ and $B$. The first case 
(Case I) is the best fit to the nuclear cross section with the fixed value 
of $A$ which is estimated by the WKB method.   They are $A=4.70\times 10^{-3}$ 
$ {\rm erg}^{1\over 2}$ and $B=17.5 \   eV$. In the second case (Case II),
 we make the best 
 fit to the nuclear cross section varying the values of the parameters 
 $A$ and $B$ freely. We find that the best fit values of $A$ and $B$ 
 are $A=4.80  \times 10^{-3}$ $ {\rm erg}^{1\over 2}$ and $B=20.0 \  eV$. 
 
As can be seen, there are obviously some ambiguities which arises 
from the difficulty of the Coulomb cross sections once we want to 
understand it to a very high accuracy. 
With these two cases of the parameters, 
we can calculate the neutrino flux in the sun. 

\vspace{2cm}

\item{\large The solar structure}

In the previous section, we have determined the parameters of the 
cross section $< \sigma v >$ for the  $^7Be (p,\gamma ) ^{8} B$ 
reaction. For other reaction cross sections, we have used the same 
values of parameters as those used in the calculation of 
Bahcall et al [15]. 

Here, we want to show our calculated result of the solar structure 
quantities. In fig.2, we show the luminosity and the temperature 
of the sun as the function of the solar radius. The solid lines are 
 the calculated results where the reaction cross section of 
 $^7Be (p,\gamma ) ^{8} B$ is used  with the parameters 
  $A$ and $B$ ( Case I ) as determined above. 
All the other nuclear data are the 
  same as those used in the calculations of Bahcall et al. 
  On the other hand, the dashed lines indicate the calculated luminosity 
  and temperature by Bahcall et al. 
  As can be seen from these figures, the 
shape of the luminosity and the temperature are almost the same 
between the two calculations.  

 Therefore, we can conclude that the solar structure 
quantities are not so much influenced by the change of nuclear reaction data of 
$^7Be (p,\gamma ) ^{8} B$, as expected. 

\vspace{2cm}

\item{\large The neutrino flux}

Since we know now how many reactions occur inside the sun, we can 
calculate the neutrino flux. 

In table 1, we show the neutrino fluxes as well as the capture rates  
at the Earth for GALLEX, KAMIOKANDE and Homestake experiments. 
In table 1a, we show the calculated results by Bahcall et al. [15] while, 
in table 1b,  the calculations by Dar and Shaviv [11] are shown. 
In table 1c, we 
show our calculated results with the Case I while, in table 1d, the results 
with the Case II are shown. 

As can be seen from the table 1,  the present calculations 
with the Case I are very similar to the ones by Dar and Shaviv. 
Therefore, it is confirmed that the KAMIOKANDE experiment is indeed 
consistent with the SSM calculations with the careful employment of the 
nuclear reaction cross section of $^7 Be (p, \gamma) ^8 B $. 

Further, the case II indicates that the ambiguity of the coulomb 
coefficient is so large that one has to be very careful for 
drawing any conclusions on the solar neutrino problems. 
At least, the result of the case II suggests that, once the $^7Be$ neutrino 
flux is suppressed, then there is a fairly good chance that all the neutrino 
experiments fall into the range of the SSM predictions.  

For the Case II, one sees that the  cross section 
of $^7 Be (p, \gamma) ^8 B $ is best fitted. Here, the Coulomb 
coefficient is slightly different from the WKB value. 
In this parameter set, we find that the neutrino flux for 
KAMIOKAMDE is a little bit too small compared to the data. 
Instead, the Homestake and GALLEX experiments will be in the range of 
the present  calculation once the $^7Be$ neutrino flux is 
suppressed. 

Also, in the Case II, the $S(0)$ value is found to be $S(0)=20 \  eV$. 
This suggests that the $S(0)$ factor depends on the factorization 
ansatz of eq.(3.3). 
 
\vspace{3cm}
\item{\large Conclusions}

We have presented a new calculation of the standard solar model 
with the emphasis on the careful 
considerations of the nuclear reactions of $^7Be (p,\gamma ) ^{8} B$.

We show here that the solar neutrino capture rates are consistent 
with the observed data for the KAMIOKANDE 
experiments. We believe that possible refinements may improve the 
accuracy of the neutrino capture rates by $20 \sim 30$ \% 
so that the GALLEX experiments may well be in the range of the SSM picture. 
In particular, the suppression of the $^7 Be$ electron capture 
inside the sun will lead to the understanding of the GALLEX and Homestake  
experiments in a natural way. 

 Therefore, we conclude that the solar neutrino fluxes are 
mostly consistent with the standard solar model with careful 
considerations of the nuclear reactions of 
$^7Be (p,\gamma ) ^{8} B$. 

In the course of the present study, we received a preprint of the new 
calculation by Dar and Shaviv which shows  very similar results 
to the present calculations. This confirms that the present result 
does not so much depend on the modeling of the sun as far as 
we take into account the gross structure of the sun. 

\end{enumerate}
\vspace{2cm}
Acknowledments: We thank C. Itoi for discussions and comments. 

\newpage 

{\Large Reference}

\vspace{1cm}

1. J.N. Bahcall and M.H. Pinsonneault, Rev. Mod. Phys. {\bf 64} (1992), 885

\quad J.N. Bahcall, Neutrino Astrophysics (Cambridge University Press, Cambridge,1989)

2. P. Mikheyev and A.Y. Smirnov, Yad. Fiz. {\bf 42} (1985), 1441 

\quad L. Wolfenstein, Phys. Rev. {\bf D17} (1978), 2369 

3. J.A. Nolen and J.P. Schiffer, Ann. Rev. Nucl. Sci. {\bf 19} (1969), 471 

4. S. Shlomo, Rep. Prog. Phys. {\bf 41} (1978), 957

5. Fillippone, B.W. et al., Phys. Rev. {\bf C 28} (1983), 2222

6. GALLEX Collaboration, P. Anselmann et al., 
Phys. Lett. {\bf B 327} (1994) 377

\quad {\bf 342} (1995) 440, \ \ {\bf 357} (1995), 237

7. K.S. Hirata et al., Phys. Rev Lett. {\bf 63} (1989), 16

\quad T. Kajita, ICRR-Report, 332-94-27 (December 1994)

8. SAGE Collaboration, S.R. Elliott et al., Nucl. Phys. {\bf B 48} (1996), 375

9. B.T. Cleveland et al., Nucl. Phys. {\bf B 38} (1995), 47 

\quad R. Davis, Prog. Part. Nucl. Phys. {\bf 32} (1994), 13

10. J. Bahcall and M. Pinsonneault, Rev. Mod. Phys. {\bf 67} (1995), 1

11. A. Dar and G. Shaviv, Ap. J. {\bf 468} (1996), 933

12. K.R. Lang, Astrophysical Formulae (Springer, Berlin, 1980)

13. B.A. Brown, A. Csoto and R. Sherr, Nucl. Phys. {\bf A597} (1996), 66

14. H.M. Xu, et al. Phys. Rev. Lett. {\bf 73} (1994), 2027

15. J.N. Bahcall and R. Ulrich, Rev. Mod. Phys. {\bf 60} (1988), 297

\newpage
{\large Table captions}

\begin{list}{}{}
\item[Table 1]: We plot the calculated neutrino flux from various 
nuclear reactions together with the experiments. Table 1a shows 
the calculation by Bahcall and Ulrich [15] while Table 1b plots 
the calculation by Dar and Shaviv [11]. Tables 1c and 1d are the present 
calculations with the parameter sets of Case I and Case II, respectively. 

\end{list} 

\vspace{2cm}

{\large Figure captions}

\begin{list}{}{}
\item[Fig.1]: 
 We show the nuclear cross section of $^7 Be (p,\gamma) ^8 B$.  
The black 
circles with error bars are the observed data by Filippone et al [5]. 
The solid line is our calculation with the Case I parameters while 
the dashed line with the Case II parameters.

\item[Fig.2]: The properties of the internal structure of the sun 
are shown as the function of the radius. The solid lines show 
the present calculations while the dashed lines the ones by Bahcall et al [15]. 
The $L$, $M$, $T$, $\rho$ and $P$ denote the luminosity, the mass, 
the temperature, the density and the pressure of the sun. 

\end{list}

\newpage
\begin{center}

\underline{Table 1a} \\
\ \ \\
\begin{tabular}{c|c|cccc}
\hline
\hline
&
 & \multicolumn{4}{c}{\raisebox{-0.2ex}[0pt]{BP88}} \\
\raisebox{0.1ex}[0pt]{source} &\raisebox{1.3ex}[0pt]{Flux}
 & \raisebox{0.2ex}[0pt]{Homestake}
 & \raisebox{0.2ex}[0pt]{GALLEX}
 & \raisebox{0.2ex}[0pt]{SAGE}
 & \raisebox{0.2ex}[0pt]{KAMIOKANDE} \\
& \raisebox{1ex}[0pt]{(cm$^{-2}$sec$^{-1}$)}
 & \raisebox{0.5ex}[0pt]{(SNU)}
 & \raisebox{0.5ex}[0pt]{(SNU)}
 & \raisebox{0.5ex}[0pt]{(SNU)}
 & \raisebox{0.5ex}[0pt]{(10$^{6}$cm$^{-2}$sec$^{-1}$)} \\
\hline
&&&& \\
\raisebox{1.3ex}[0pt]{$pp$}
 & \raisebox{1.3ex}[0pt]{$6.0\times10^{10}$}
 & \raisebox{1.3ex}[0pt]{---}
 & \raisebox{1.3ex}[0pt]{70.8}
 & \raisebox{1.3ex}[0pt]{70.8}
 & \raisebox{1.3ex}[0pt]{---} \\
\raisebox{1.3ex}[0pt]{$pep$}
 & \raisebox{1.3ex}[0pt]{$1.4\times10^{8}$}
 & \raisebox{1.3ex}[0pt]{0.2}
 & \raisebox{1.3ex}[0pt]{ 3.0}
 & \raisebox{1.3ex}[0pt]{ 3.0}
 & \raisebox{1.3ex}[0pt]{---} \\
\raisebox{1.3ex}[0pt]{$^{7}$Be}
 & \raisebox{1.3ex}[0pt]{$4.7\times10^{9}$}
 & \raisebox{1.3ex}[0pt]{1.1}
 & \raisebox{1.3ex}[0pt]{34.3}
 & \raisebox{1.3ex}[0pt]{34.3}
 & \raisebox{1.3ex}[0pt]{---} \\
\raisebox{1.3ex}[0pt]{$^{8}$B}
 & \raisebox{1.3ex}[0pt]{$5.8\times10^{6}$}
 & \raisebox{1.3ex}[0pt]{6.1}
 & \raisebox{1.3ex}[0pt]{14.0}
 & \raisebox{1.3ex}[0pt]{14.0}
 & \raisebox{1.3ex}[0pt]{5.8} \\
\raisebox{1.3ex}[0pt]{$^{13}$N}
 & \raisebox{1.3ex}[0pt]{$6.1\times10^{8}$}
 & \raisebox{1.3ex}[0pt]{0.1}
 & \raisebox{1.3ex}[0pt]{ 3.8}
 & \raisebox{1.3ex}[0pt]{ 3.8}
 & \raisebox{1.3ex}[0pt]{---} \\
\raisebox{1.3ex}[0pt]{$^{15}$O}
 & \raisebox{1.3ex}[0pt]{$5.2\times10^{8}$}
 & \raisebox{1.3ex}[0pt]{0.3}
 & \raisebox{1.3ex}[0pt]{ 6.1}
 & \raisebox{1.3ex}[0pt]{ 6.1}
 & \raisebox{1.3ex}[0pt]{---} \\
\hline
&&&& \\
\raisebox{1.3ex}[0pt]{Total} && \raisebox{1.3ex}[0pt]{7.9}
 & \raisebox{1.3ex}[0pt]{132}
 & \raisebox{1.3ex}[0pt]{132}
 & \raisebox{1.3ex}[0pt]{5.8} \\
\hline
\hline
&&&& \\
\raisebox{1.3ex}[0pt]{Experiment} && \raisebox{1.3ex}[0pt]{$2.55\pm0.25$}
 & \raisebox{1.3ex}[0pt]{$77.1\pm8.5^{+4.4}_{-5.4}$}
 & \raisebox{1.3ex}[0pt]{$69\pm10^{+5}_{-7}$}
 & \raisebox{1.3ex}[0pt]{$2.73\pm0.17\pm0.34$} \\
\hline
\hline
\end{tabular}

\ \ \\
\ \ \\
\ \ \\

\underline{Table 1b} \\
\ \ \\
\begin{tabular}{c|c|cccc}
\hline
\hline
&
 & \multicolumn{4}{c}{\raisebox{-0.2ex}[0pt]{DS96}} \\
\raisebox{0.1ex}[0pt]{source} &\raisebox{1.3ex}[0pt]{Flux}
 & \raisebox{0.2ex}[0pt]{Homestake}
 & \raisebox{0.2ex}[0pt]{GALLEX}
 & \raisebox{0.2ex}[0pt]{SAGE}
 & \raisebox{0.2ex}[0pt]{KAMIOKANDE} \\
& \raisebox{1ex}[0pt]{(cm$^{-2}$sec$^{-1}$)}
 & \raisebox{0.5ex}[0pt]{(SNU)}
 & \raisebox{0.5ex}[0pt]{(SNU)}
 & \raisebox{0.5ex}[0pt]{(SNU)}
 & \raisebox{0.5ex}[0pt]{(10$^{6}$cm$^{-2}$sec$^{-1}$)} \\
\hline
&&&& \\
\raisebox{1.3ex}[0pt]{$pp$}
 & \raisebox{1.3ex}[0pt]{$6.1\times10^{10}$}
 & \raisebox{1.3ex}[0pt]{---}
 & \raisebox{1.3ex}[0pt]{72.0}
 & \raisebox{1.3ex}[0pt]{72.0}
 & \raisebox{1.3ex}[0pt]{---} \\
\raisebox{1.3ex}[0pt]{$pep$}
 & \raisebox{1.3ex}[0pt]{$1.43\times10^{8}$}
 & \raisebox{1.3ex}[0pt]{0.20}
 & \raisebox{1.3ex}[0pt]{3.06}
 & \raisebox{1.3ex}[0pt]{3.06}
 & \raisebox{1.3ex}[0pt]{---} \\
\raisebox{1.3ex}[0pt]{$^{7}$Be}
 & \raisebox{1.3ex}[0pt]{$3.71\times10^{9}$}
 & \raisebox{1.3ex}[0pt]{0.87}
 & \raisebox{1.3ex}[0pt]{27.1}
 & \raisebox{1.3ex}[0pt]{27.1}
 & \raisebox{1.3ex}[0pt]{---} \\
\raisebox{1.3ex}[0pt]{$^{8}$B}
 & \raisebox{1.3ex}[0pt]{$2.49\times10^{6}$}
 & \raisebox{1.3ex}[0pt]{2.62}
 & \raisebox{1.3ex}[0pt]{6.01}
 & \raisebox{1.3ex}[0pt]{6.01}
 & \raisebox{1.3ex}[0pt]{2.49} \\
\raisebox{1.3ex}[0pt]{$^{13}$N}
 & \raisebox{1.3ex}[0pt]{$3.82\times10^{8}$}
 & \raisebox{1.3ex}[0pt]{0.06}
 & \raisebox{1.3ex}[0pt]{2.38}
 & \raisebox{1.3ex}[0pt]{2.38}
 & \raisebox{1.3ex}[0pt]{---} \\
\raisebox{1.3ex}[0pt]{$^{15}$O}
 & \raisebox{1.3ex}[0pt]{$3.74\times10^{8}$}
 & \raisebox{1.3ex}[0pt]{0.22}
 & \raisebox{1.3ex}[0pt]{4.39}
 & \raisebox{1.3ex}[0pt]{4.39}
 & \raisebox{1.3ex}[0pt]{---} \\
\hline
&&&& \\
\raisebox{1.3ex}[0pt]{Total} && \raisebox{1.3ex}[0pt]{4.1}
 & \raisebox{1.3ex}[0pt]{115}
 & \raisebox{1.3ex}[0pt]{115}
 & \raisebox{1.3ex}[0pt]{2.49} \\
\hline
\hline
&&&& \\
\raisebox{1.3ex}[0pt]{Experiment} && \raisebox{1.3ex}[0pt]{$2.55\pm0.25$}
 & \raisebox{1.3ex}[0pt]{$77.1\pm8.5^{+4.4}_{-5.4}$}
 & \raisebox{1.3ex}[0pt]{$69\pm10^{+5}_{-7}$}
 & \raisebox{1.3ex}[0pt]{$2.73\pm0.17\pm0.34$} \\
\hline
\hline
\end{tabular}

\newpage
\underline{Table 1c} \\
\ \ \\
\begin{tabular}{c|c|cccc}
\hline
\hline
&
 & \multicolumn{4}{c}{\raisebox{-0.2ex}[0pt]{Present cal.I}} \\
\raisebox{0.1ex}[0pt]{source} &\raisebox{1.3ex}[0pt]{Flux}
 & \raisebox{0.2ex}[0pt]{Homestake}
 & \raisebox{0.2ex}[0pt]{GALLEX}
 & \raisebox{0.2ex}[0pt]{SAGE}
 & \raisebox{0.2ex}[0pt]{KAMIOKANDE} \\
& \raisebox{1ex}[0pt]{(cm$^{-2}$sec$^{-1}$)}
 & \raisebox{0.5ex}[0pt]{(SNU)}
 & \raisebox{0.5ex}[0pt]{(SNU)}
 & \raisebox{0.5ex}[0pt]{(SNU)}
 & \raisebox{0.5ex}[0pt]{(10$^{6}$cm$^{-2}$sec$^{-1}$)} \\
\hline
&&&& \\
\raisebox{1.3ex}[0pt]{$pp$}
 & \raisebox{1.3ex}[0pt]{$5.7\times10^{10}$}
 & \raisebox{1.3ex}[0pt]{---}
 & \raisebox{1.3ex}[0pt]{67.3}
 & \raisebox{1.3ex}[0pt]{67.3}
 & \raisebox{1.3ex}[0pt]{---} \\
\raisebox{1.3ex}[0pt]{$pep$}
 & \raisebox{1.3ex}[0pt]{$1.4\times10^{8}$}
 & \raisebox{1.3ex}[0pt]{0.20}
 & \raisebox{1.3ex}[0pt]{3.00}
 & \raisebox{1.3ex}[0pt]{3.00}
 & \raisebox{1.3ex}[0pt]{---} \\
\raisebox{1.3ex}[0pt]{$^{7}$Be}
 & \raisebox{1.3ex}[0pt]{$4.7\times10^{9}$}
 & \raisebox{1.3ex}[0pt]{1.10}
 & \raisebox{1.3ex}[0pt]{34.3}
 & \raisebox{1.3ex}[0pt]{34.3}
 & \raisebox{1.3ex}[0pt]{---} \\
\raisebox{1.3ex}[0pt]{$^{8}$B}
 & \raisebox{1.3ex}[0pt]{$2.9\times10^{6}$}
 & \raisebox{1.3ex}[0pt]{3.05}
 & \raisebox{1.3ex}[0pt]{7.00}
 & \raisebox{1.3ex}[0pt]{7.00}
 & \raisebox{1.3ex}[0pt]{2.9} \\
\raisebox{1.3ex}[0pt]{$^{13}$N}
 & \raisebox{1.3ex}[0pt]{$3.7\times10^{8}$}
 & \raisebox{1.3ex}[0pt]{0.06}
 & \raisebox{1.3ex}[0pt]{2.30}
 & \raisebox{1.3ex}[0pt]{2.30}
 & \raisebox{1.3ex}[0pt]{---} \\
\raisebox{1.3ex}[0pt]{$^{15}$O}
 & \raisebox{1.3ex}[0pt]{$2.2\times10^{8}$}
 & \raisebox{1.3ex}[0pt]{0.13}
 & \raisebox{1.3ex}[0pt]{2.58}
 & \raisebox{1.3ex}[0pt]{2.58}
 & \raisebox{1.3ex}[0pt]{---} \\
\hline
&&&& \\
\raisebox{1.3ex}[0pt]{Total} && \raisebox{1.3ex}[0pt]{4.5}
 & \raisebox{1.3ex}[0pt]{116}
 & \raisebox{1.3ex}[0pt]{116}
 & \raisebox{1.3ex}[0pt]{2.9} \\
\hline
\hline
&&&& \\
\raisebox{1.3ex}[0pt]{Experiment} && \raisebox{1.3ex}[0pt]{$2.55\pm0.25$}
 & \raisebox{1.3ex}[0pt]{$77.1\pm8.5^{+4.4}_{-5.4}$}
 & \raisebox{1.3ex}[0pt]{$69\pm10^{+5}_{-7}$}
 & \raisebox{1.3ex}[0pt]{$2.73\pm0.17\pm0.34$} \\
\hline
\hline
\end{tabular}

\ \ \\
\ \ \\
\ \ \\

\underline{Table 1d} \\
\ \ \\
\begin{tabular}{c|c|cccc}
\hline
\hline
&
 & \multicolumn{4}{c}{\raisebox{-0.2ex}[0pt]{Present cal.II}} \\
\raisebox{0.1ex}[0pt]{source} &\raisebox{1.3ex}[0pt]{Flux}
 & \raisebox{0.2ex}[0pt]{Homestake}
 & \raisebox{0.2ex}[0pt]{GALLEX}
 & \raisebox{0.2ex}[0pt]{SAGE}
 & \raisebox{0.2ex}[0pt]{KAMIOKANDE} \\
& \raisebox{1ex}[0pt]{(cm$^{-2}$sec$^{-1}$)}
 & \raisebox{0.5ex}[0pt]{(SNU)}
 & \raisebox{0.5ex}[0pt]{(SNU)}
 & \raisebox{0.5ex}[0pt]{(SNU)}
 & \raisebox{0.5ex}[0pt]{(10$^{6}$cm$^{-2}$sec$^{-1}$)} \\
\hline
&&&& \\
\raisebox{1.3ex}[0pt]{$pp$}
 & \raisebox{1.3ex}[0pt]{$5.7\times10^{10}$}
 & \raisebox{1.3ex}[0pt]{---}
 & \raisebox{1.3ex}[0pt]{67.3}
 & \raisebox{1.3ex}[0pt]{67.3}
 & \raisebox{1.3ex}[0pt]{---} \\
\raisebox{1.3ex}[0pt]{$pep$}
 & \raisebox{1.3ex}[0pt]{$1.4\times10^{8}$}
 & \raisebox{1.3ex}[0pt]{0.20}
 & \raisebox{1.3ex}[0pt]{3.00}
 & \raisebox{1.3ex}[0pt]{3.00}
 & \raisebox{1.3ex}[0pt]{---} \\
\raisebox{1.3ex}[0pt]{$^{7}$Be}
 & \raisebox{1.3ex}[0pt]{$4.7\times10^{9}$}
 & \raisebox{1.3ex}[0pt]{1.10}
 & \raisebox{1.3ex}[0pt]{34.3}
 & \raisebox{1.3ex}[0pt]{34.3}
 & \raisebox{1.3ex}[0pt]{---} \\
\raisebox{1.3ex}[0pt]{$^{8}$B}
 & \raisebox{1.3ex}[0pt]{$1.9\times10^{6}$}
 & \raisebox{1.3ex}[0pt]{1.95}
 & \raisebox{1.3ex}[0pt]{4.47}
 & \raisebox{1.3ex}[0pt]{4.47}
 & \raisebox{1.3ex}[0pt]{1.9} \\
\raisebox{1.3ex}[0pt]{$^{13}$N}
 & \raisebox{1.3ex}[0pt]{$3.7\times10^{8}$}
 & \raisebox{1.3ex}[0pt]{0.06}
 & \raisebox{1.3ex}[0pt]{2.30}
 & \raisebox{1.3ex}[0pt]{2.30}
 & \raisebox{1.3ex}[0pt]{---} \\
\raisebox{1.3ex}[0pt]{$^{15}$O}
 & \raisebox{1.3ex}[0pt]{$2.2\times10^{8}$}
 & \raisebox{1.3ex}[0pt]{0.13}
 & \raisebox{1.3ex}[0pt]{2.58}
 & \raisebox{1.3ex}[0pt]{2.58}
 & \raisebox{1.3ex}[0pt]{---} \\
\hline
&&&& \\
\raisebox{1.3ex}[0pt]{Total} && \raisebox{1.3ex}[0pt]{3.4}
 & \raisebox{1.3ex}[0pt]{114}
 & \raisebox{1.3ex}[0pt]{114}
 & \raisebox{1.3ex}[0pt]{1.9} \\
\hline
\hline
&&&& \\
\raisebox{1.3ex}[0pt]{Experiment} && \raisebox{1.3ex}[0pt]{$2.55\pm0.25$}
 & \raisebox{1.3ex}[0pt]{$77.1\pm8.5^{+4.4}_{-5.4}$}
 & \raisebox{1.3ex}[0pt]{$69\pm10^{+5}_{-7}$}
 & \raisebox{1.3ex}[0pt]{$2.73\pm0.17\pm0.34$} \\
\hline
\hline
\end{tabular}

\end{center}

\end{document}